\documentclass{article}
\usepackage{geometry}
\usepackage[utf8]{inputenc}
\usepackage[american]{babel}
\usepackage{authblk}
\usepackage[autostyle,autopunct=true]{csquotes}
\usepackage{graphicx}
\usepackage{amsmath}
\usepackage{amssymb}
\usepackage{amsthm}
\usepackage[
    colorlinks=true,
    linkcolor=blue,
    citecolor=blue,
    urlcolor=blue
]{hyperref}
\usepackage[
    backend=biber,
    style=apa,
    citestyle=authoryear,
    natbib=true
]{biblatex}
\usepackage{floatrow}
\usepackage[
    font={sf},
    justification=raggedright,
    singlelinecheck=off,
    labelsep=period
]{caption}
\usepackage{threeparttable}

\title{Communication networks and group effectiveness:\\the case of English Wikipedia}
\author[1]{%
    Agnieszka Rychwalska
    \thanks{\href{mailto:a.rychwalska@uw.edu.pl}{\texttt{a.rychwalska@uw.edu.pl}}}
}
\author[1]{Szymon Talaga}
\author[3]{Karolina Ziembowicz}
\author[4]{Dariusz Jemielniak}
\affil[1]{Robert Zajonc Institute for Social Studies, University of Warsaw, Poland}
\affil[2]{Maria Grzegorzewska University, Poland}
\affil[3]{Kozminski University, Poland and Berkman-Klein Center for Internet and Society, Harvard University, US}

\graphicspath{figures/}

\newcommand{\Det}{\ensuremath{\text{Det}}}
\newcommand{\Deg}{\ensuremath{\text{Deg}}}
\newcommand{\EI}{\ensuremath{\text{EI}}}

\begin{document}

\maketitle

\begin{abstract}
    A selection of intellectual goods produced by online communities – e.g. open source software or knowledge bases like Wikipedia – are in daily use by a broad audience, and thus their quality impacts the public at large. Yet, it is still unclear what contributes to the effectiveness of such online peer production systems: what conditions or social processes help them deliver quality products. Specifically, while co-contribution (i.e. bipartite networks) are often investigated in online collaboration, the role of interpersonal communication in coordination of online peer-production is much less investigated. To address this gap we have reconstructed networks of personal communication (direct messaging) between Wikipedia editors gathered in so called Wikiprojects – teams of contributors who focus on articles within specific topical areas. We found that effective projects exchange larger volume of direct messages and that their communication structure allows for complex coordination: for sharing of information locally through selective ties, and at the same time globally across the whole group. To verify how these network measures relate to the subjective perception of importance of group communication we conducted semi-structured interviews with members of selected projects. Our interviewees used direct communication for providing feedback, for maintaining close relations and for tapping on the social capital of the Wikipedia community. Our results underscore the importance of communication structure in online collaboration: online peer production communities rely on interpersonal communication to coordinate their work and to maintain high levels of engagement. Design of platforms for such communities should allow for ample group level communication as well as for one-on-one interactions.
\end{abstract}

\section{Introduction}

Arguably, individuals form groups to achieve more ambitious goals than they could achieve alone~\citep{allportInfluenceGroupAssociation1920,caporaelPartsWholesEvolutionary2001}. Yet, for groups to proverbially become more than the sum of their parts, they need to organize into effective structures: division of labor~\cite{rozasLoosenControlLosing2021}, structures of information flow and the structure of interpersonal relations~\citep{akdereEconomicsSocialCapital2008}, as well as a well internalized organizational culture are crucial for group performance.

Not surprisingly then, the structure of online collaboration teams and peer production systems, such as Free/Open Source Software (FOSS) or Wikipedia, have often been analyzed to pinpoint those structural properties that determine the quality of group work. Bipartite, collaboration networks (e.g.~co-editing of Wikipedia articles or co-contributions to files in FOSS projects) are the most likely targets of such analysis, as they show how contributors interact with each other when performing the actual tasks. For example, longer path lengths in co-edition networks of groups of editors on English Wikipedia are related to lower performance of the group~\citep{plattNetworkStructureEfficiency2018}. In FOSS projects, contributors self-organize into subgroups working on similar subsets of files in a bottom-up manner~\citep{palazziOnlineDivisionLabour2019}, by performing spontaneously self-chosen and self-defined tasks~\citep{chelkowskiInequalitiesOpenSource2016}, or self-select to tasks (e.g.~modules) in a well-defined hierarchy of labor division~\citep{mozillaRolesLeadership2021Special}.

However, a long-standing line of research on group work in offline settings shows that task-related structuring of the group is not enough for high performance. Task-related group interaction is accompanied by affiliation-related interaction~\citep{balesEquilibriumProblemSmall1953,balesHowPeopleInteract1955}. The task (instrumental) domain of group dynamics drives the group to differentiate, while the affiliation processes (affective domain) focus on integration and promote uniformity~\citep{balesHowPeopleInteract1955}.

Affiliation processes are important not only for the wellbeing of group members: group cohesion is positively related to performance~\citep{bealCohesionPerformanceGroups2003,evansGroupCohesionPerformance1991}. While instrumental, task-related activities can be organized by procedures and rules, social bonding within the group crucially relies on interpersonal communication. In organizations, coordination of work often happens in informal communication, alongside formal structures of management~\citep{beerDecisionControlMeaning1994}. Thus, the structure of interpersonal communication within online peer production groups - specifically, its cohesiveness - should affect the groups’ performance. 

In this work, we focus on communication using direct messages within collaborating groups on the English Wikipedia. We used a mixed-methods approach - quantitative social network analysis and qualitative analysis of interviews - to analyze Wikipedia’s topic oriented editor groups: Wikiprojects. By reconstructing networks of direct messages and interviewing core members from selected projects we were able to elucidate how interpersonal interaction, and specifically affiliation related communication, impacts quality of group work. We show that high quality of group work is related to communication networks with strong ties, which also exhibit integrated connection structure, straying off the typical leader – followers, star-like motifs. We also found that direct messaging is associated with motivational messages and outreach, and with tapping on the group’s social capital. Taken together, our results suggest that communication structures that allow for complex integration of resources and knowledge distributed across specialized contributors within the group are conducive to group efficiency.

\section{Background and related work}

Quality of Wikipedia has been a topic in both public and academic discourse since its inception~\citep{jemielniakCulturalDiversityQuality2017,shafeeEvolutionWikipediaMedical2017,smith2020situating}. On the one hand, its popularity combined with lack of professional oversight over the content, raised concerns of possible misinformation of the general public~\citep{lanierDigitalMaoismHazards2006}. On the other hand, the process of bottom up encyclopedic content creation raised interest in self-organization possibilities afforded by new media~\citep{shirkyHereComesEverybody2008}. Thus, understanding how quality products emerge from bottom up social processes is both important for the public at large as well as for theoretical advancement of systems research.

With the growth of Wikipedia, and with refinement of its peer production process, the understanding of self-organized quality control has been accumulating~\citep{nemotoSocialCapitalIncreases2011,plattNetworkStructureEfficiency2018,qinInfluenceNetworkStructures2015}. Most importantly, Wikipedia’s peer review process and quality scores awarded to articles provide an opportunity to quantify the effects of distributed work and allow comparisons against a single scale of quality ratings. 

\subsection{Quality in peer production}

For the average reader of Wikipedia, the quality grades of Featured Article (thereafter referred to as FA) or Good Article (GA) – a star or a plus sign, respectively, displayed at the top of the page – might be unnoticeable, but for Wikipedia editors they are cherished marks of accomplishment. Many editors list the FA and GAs they contributed to on their Wikipedia profile pages. The review process that awards those ratings is in many respects similar to academic reviews: the criteria for each rank of article quality are listed explicitly, and each submission is reviewed by independent editors whose remarks are then incorporated into the article. The difference lies in that at the end of the process, a discussion unfolds, and any editor can voice their opinion on whether the article is worthy of a particular quality mark; the ensuing decision is consensus based\footnote{%
    \url{https://en.wikipedia.org/wiki/Wikipedia:Featured_article_review}
}.%
The universal criteria (well-known and applicable to all candidate articles) as well as independent review process makes these quality ratings suitable for a quality measurement of peer produced work.

The availability of such quality measures has spurred much investigation into what constitutes quality work. Interestingly, quality criteria differ among language versions of Wikipedia, pointing to possible differences in quality perceptions across cultures~\citep{jemielniakCulturalDiversityQuality2017}. On one hand, particular features of articles that predict well how close they are to being awarded a higher quality grade, can be identified. For example, classifiers trained on previously assessed articles can tell how close an article is to a certain quality rank~\citep{warncke2013tell}. On the other hand, the process of article production can also be investigated to pinpoint those properties that enable some articles to improve in quality much faster than others.

In standard organizations’ production process, quality is ensured, among others, through a careful division of labor among employees with appropriate expertise. In online peer production systems, contributors self-select to tasks~\citep{benklerCoasePenguinLinux2002}, and thus it has been suggested that composition of the editor base impacts quality of Wikipedia articles. \citet{sydow2017diversity} have found that articles edited by contributors of rich and broad background – i.e. those that in their history have edited many varied articles – are more likely to reach FA status. This might not only be an effect of broader subject expertise of these editors but also of a process of \enquote{creative abrasion} wherein editors with varied knowledge and experience engage in discussions to resolve their differences, which leads to article improvement~\citep{arazyInformationQualityWikipedia2011}. Whilst some abrasion might be positive and promote quality, too much of conflict can also be detrimental~\citep{arazyStayWikipediaTask2013}. It seems there is an optimal level of diversity of contributors working on an article and that it is moderated by the size of the group: smaller groups can be less diverse and still deliver high quality articles~\citep{robertCrowdSizeDiversity2015}.

Diversity might be a factor positively impacting quality, but combining diverse inputs requires processes of coordination, which are recognized as a crucial aspect of quality control in standard organizations~\citep{beerViableSystemModel1984}. In online social systems coordination is peer produced in a similar manner to product development~\citep{benklerCoasePenguinLinux2002}. That is, contributors self-select to perform coordination tasks. Not surprisingly, coordination plays a part in creating quality articles on Wikipedia: even such simple coordination as splitting into core and periphery roles (defined as editors who submit the bulk of content vs. editors who make minor additions) increases the chances for FA status~\citep{kitturHarnessingWisdomCrowds2008}. Interestingly, without such a split, simple increase in the number of editors did not lead to improved quality (ibid.).

Finally, an important factor impacting quality is engagement. In standard organizations motivation and conscientiousness can be promoted by various incentives, including financial ones. In online peer production other incentives are at play: ranging from reputation, to self-development, to socializing~\citep{oregExploringMotivationsContributing2008}. Engagement can be induced or sustained by interactions with other, well-socialized members of the community: welcoming messages, assistance and constructive criticism increase commitment of new editors on Wikipedia and increase the chances for their prolonged contribution~\citep{choiSocializationTacticsWikipedia2010}. Such motivational activities are mostly carried out by sending direct messages, i.e. editing a newcomer’s personal talk page~\citep{zhu2011identifying}. Whilst core community members send more such messages per person, in total peripheral members deliver the bulk of engagement related direct messaging, contributing to the shared leadership within Wikipedia~(ibid.). Interestingly, within the thematic oriented groups on Wikipedia (Wikiprojects) task related and feedback related direct messages are initiated more often by peripheral members while socializing aimed messages are sent more often by core members~(ibid.). This suggests that just as it happens in offline groups~\citep{balesHowPeopleInteract1955}, in online peer production communities social and task leadership is divided to a degree.

\subsection{Direct communication in group work}

Interpersonal, one-on-one communication not only affects motivation of group members but it is also the vehicle for much coordination in standard organizations. Whilst formal oversight and management are important for policy and top-level management, coordination of tasks is often carried out in informal meetings or communication~\citep{beerDecisionControlMeaning1994}. Dyadic relationships and communication play a role in maintaining motivation, engagement and cohesiveness of a group~\citep{lidenDyadicRelationships2016}, contributing to the affective domain within the group~\citep{balesHowPeopleInteract1955}. Through dyadic interactions, members of the group transmit emotional support and share task-related resources~\citep{omilion-hodgesContextualizingLMXWorkgroup2013}. Direct communication also warrants social learning, i.e. learning that occurs through observation and direct instruction of more experienced team members~\citep{singh2013social}.

In digital collaboration, personal messages have also been identified as impacting group work. For example, informal, direct communication among distributed workers helps to build bonds and ensures effective information flow~\citep{hindsUnderstandingConflictGeographically2005}. The importance of direct communication in distributed teams is corroborated by the fact that activity on user pages on the English Wikipedia continued to grow in spite of declining number of edits on article pages~\citep{nemotoSocialCapitalIncreases2011}. Moreover, disassortative direct communication network structures – i.e.~linkages between hubs and periphery of the networks – are prevalent among Wikipedia’s Wikiproject teams and mild disassortativity is related to higher quality of articles that these projects curate~\citep{rychwalskaQualityPeerProduction2020}.

\subsection{Summary and hypotheses}

The effectiveness of an online collaboration group – i.e.~the quality of its products – can~thus be related to both its task related structuration that ensures proper task performance and to interpersonal relations (affiliation processes) that allow coordination and promote engagement~\citep{balesEquilibriumProblemSmall1953}. The former has been extensively studied in both Wikipedia and FOSS by investigating division of labor and (co-)contribution patterns (e.g.~\cite{arazyFunctionalRolesCareer2015,plattNetworkStructureEfficiency2018}). Relation building and motivational structures, on the other hand, have been much less investigated; yet offline theory and research of group processes as well as initial investigations into the role of personal messaging indicate that it plays a crucial role in group effectiveness. Specifically, the content of direct communication has been shown to impact motivation to contribute~\citep{choiSocializationTacticsWikipedia2010} and to relate to relational leadership~\citep{zhu2011identifying}, however the structure of such communication may also play a role.

On Wikipedia this structure can be gleaned from direct messages sent between editors and analyzed as a network~\citep{rychwalskaQualityPeerProduction2020}. If indeed direct messages are a vehicle for building and maintaining engagement, 
\textbf{(H1)} successful collaborating groups will exchange a larger volume of such messages than unsuccessful ones. Moreover, if they are a means for self-regulation and coordination 
\textbf{(H2)}, direct communication structure of successful teams will allow for both local, specialized information sharing as well as global integration of information within the group~\citep{kleinEmergenceInformativeHigher2020}. Finally, if they serve as a vehicle for group integration
\textbf{(H3)} direct messages will link a larger part of the group (rather than some selected group members) within successful groups than unsuccessful ones. To test these three hypotheses we have designed an exploratory study into the communication structure of thematic groups on English Wikipedia: Wikiprojects, and into its impact on the effects of group work: quality of articles. For each project we have reconstructed a network of direct messaging and measured the related properties: average node strength, network degeneracy and determinism (components of effective information as defined by~\citet{kleinEmergenceInformativeHigher2020}), and fraction of project editors that take part in direct communication network. We related these network properties to the quality of articles within the relevant project’s scope and we found that direct communication networks with high effective information and strong links are related to more effective group work. To investigate in depth the role of such cohesive communication networks in Wikiprojects’ work and with results of our quantitative analysis in mind, we designed semi-structured interviews that we then carried out with selected members of two projects: the one that was most successful with regard to producing quality articles (Tropical Cyclones), and one that was the fastest growing project in 2020 (COVID-19). We found that direct messaging is crucial for feedback on individual work and that one-on-one communication networks play a part in tapping on the groups’ social capital.

\section{Methods}

We conducted a Thick Big Data analysis~\citep{jemielniakThickBigData2020} of collaboration on English Wikipedia. Thick Big Data is a novel mixed-method approach, allowing for combining quantitative analysis of large datasets with deep, qualitative inquiry. In our case, the analysis relied on combining qualitative interviews with heavily engaged Wikiproject participants and network analysis of direct communication within the projects.

\subsection{Wikiprojects}

\begin{figure}[!t]
    \centering
    \includegraphics[width=\textwidth]{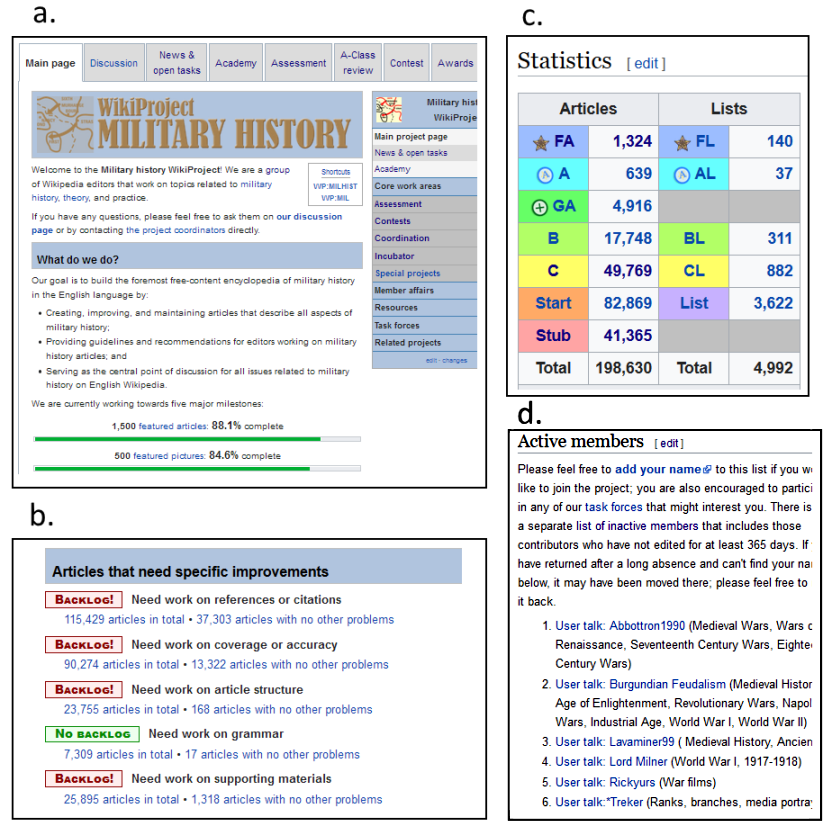}
    \caption{Screenshots from the Wikipedia coordination space for Wikiproject Military History: (a.) the front page, delineating the goal of the project; (b.) fragment of a section with pending tasks for members to work on; (c.) quality statistics for the project; and (d.) fragment of the member list where Wikipedians can join the project.}
    \label{fig:wp-diagram}
\end{figure}

As our unit of analysis we have chosen English Wikipedia’s Wikiprojects. These are groups of editors that work together on encyclopedic articles related to a specific topic\footnote{%
    \url{https://en.wikipedia.org/wiki/Wikipedia:WikiProject}
}.%
For example, there is a project on Military History, on Michael Jackson or on the Simpsons. Projects vary in scope (i.e.~the number of articles they curate), depending both on the breadth of the subject as well as on its popularity among Wikipedians (the more editors are interested in a particular topic, the more articles they will create and the better the coverage of this topic on Wikipedia). There is one thing all projects share, though: they all strive to improve the quality of articles within their scope. To achieve this goal they use various methods of coordination - from discussion on article importance, to editathons and drives to improve articles, to templates and preparing manuals of style, among others. Each Wikiproject has a dedicated space on the so-called project namespace on Wikipedia. In there, the goals, rules and procedures, as well as current information on the project are displayed~(Fig.~\ref{fig:wp-diagram}). These spaces also include talk pages that are used for discussion of current issues within the project.

\subsection{Data gathering and network construction}

Source code of all user talk pages on English Wikipedia was collected via Wikipedia API\footnote{%
    \url{https://en.wikipedia.org/w/api.php}
}% 
during the first half of February 2021 (the process ended on February 11) using the \textit{CirrusDoc} query endpoint. Simultaneously, information on all quality assessments of articles available in the main encyclopedia (so-called main namespace) was extracted using the \textit{AllPages} query method of Wikipedia API (the process ended on February 12). Inconsistent naming of Wikiprojects was standardized. The extraction process spanned several days as in both cases a high level of auto-throttling was used in order not to consume too much of Wikipedia API bandwidth.

Next, we defined sets of Wikipedia editors associated with individual projects as all editors who left at least one signed message on one of the project pages (but not talk pages) belonging to a given Wikiproject. Crucially, project pages also include lists of project members which consist of (usually) user signatures. Hence, we included all editors properly signed on member lists as well as all who contributed in a detectable way to creation of project pages of a given Wikiproject. This second case is important as not all editors active within a project choose to sign up to member lists. We consciously did not include all editors who contributed to the encyclopedic articles curated by projects, as such lists would include many accidental editors, not collaborating within the projects.

Then, we used a discussion parser tailored for Wikipedia talk pages~\citep{talagaSztalWikitalkparserInitial2021} to extract all discussion threads from talk pages of all users who were identified as Wikiproject editors. The threads selected by the parser included template based messages (such as warnings, barnstars, welcome messages, etc.) but did not include mass messaging, such as newsletters. 

Finally, the discussion threads were used to define undirected weighted links between pairs of editors who left messages on each other’s user pages with weights corresponding to the number of interactions between them. We chose to use undirected links as it is usually very hard to keep track of messages and their replies due to many different communication customs used on Wikipedia. For instance, in some cases an original question left on a user’s talk page may be answered by the user on that exact same page while in other cases the reply may be left on the talk page of the editor asking the original question. This is why we prefer to consider any message between two users as an instance of an undirected interaction between them.

We used the above approach to recreate a total of 1625 networks. However, for downstream analyses we used only a subset of projects with at least 5 editors being present in the direct communication network. This filtering step was necessary as most structural network metrics are not really meaningful for smaller networks as they are constrained to take values from only a very small and discrete set. Moreover, projects with no FA/GA pages were also excluded: having produced no quality output these projects may well comprise a different population and hence would not help us address our research questions on impact of interpersonal communication on group efficiency. After these two filtering steps the initial set of 1625 direct communication networks was reduced to 997 observations of undirected networks.

\subsection{Interviews: choice of respondents and procedure}

For the purpose of our qualitative part, we conducted four interviews: two with early participants from Wikiproject Covid-19, and two with early participants from Wikiproject Tropical Cyclones. We reached out to eight of the early editors listed on the Wikiprojects’ pages. We focused on those who are still active and who allow contact by email. Four people finally agreed to an interview. The interviews took about one hour each, and were conducted via Google Meet platform (with video). All interviews were recorded with consent and transcribed. The interviews were loosely structured and open-ended~\citep{ciesielskaQualitativeMethodologiesOrganization2018}. The questions were grouped into a few themes that reflected our hypotheses: engagement, project coordination, and communication within the project. The sample questions included:
\begin{itemize}
    \item \textbf{Engagement and outreach}
    \begin{itemize}
        \item How did the project start?
        \item What makes a successful wikiproject?
        \item How are the new members recruited and on-boarded? 
    \end{itemize}
    \item \textbf{Project coordination, division of tasks and governance}
    \begin{itemize}
        \item Why do you participate, and how do you choose what to do? 
        \item What does the leadership of the project look like? 
        \item What could you tell about the changes in coordination of the project over the years?
    \end{itemize}
    \item \textbf{Communication}
    \begin{itemize}
        \item How do project members communicate? 
    \end{itemize}
\end{itemize}

\begin{table}[h]
    \centering
    \caption{Interviewee list}
    \begin{tabular}{lll}
        User & Active since & Project  \\
        \hline
        Interviewee 1 & 2007 & COVID-19 \\
        Interviewee 2 & 2006 & Tropical Cyclones \\
        Interviewee 3 & 2005 & Tropical Cyclones \\
        Interviewee 4 & 2003 & COVID-19 \\
        \hline
    \end{tabular}
    \label{tab:interviewees}
\end{table}

\section{Results – network analysis}

\subsection{Analytical strategy and measures}

Our hypotheses are centered on the impact of communication structure on the capacities of a group to deliver high quality products by collaboration. By reviewing extant findings on quality work in peer production we pinpointed that direct, one-on-one communication can bring in group integration, engagement and complex coordination patterns, all required to produce complex artifacts of high quality such as featured Wikipedia articles. However, these theoretical concepts can be measured in multiple ways and for the results to be meaningful a lot depends on proper operationalization.

\subsection{Group integration}

Following the results of~\citet{zhu2011identifying} we assume that both peripheral and core members are involved in leadership behaviors aimed at affiliation processes. Thus an important aspect of quality Wikiproject work would be to encourage communication between as many members of the group as possible. This can be measured as the fraction of project members that are included in the direct communication network.

\subsection{Engagement-building strong ties}

Following the work of~\citet{balesHowPeopleInteract1955} we assume that affective processes within a group are important for group effectiveness. Moreover we accept, after~\citet{oregExploringMotivationsContributing2008} that engagement of group members results in sustained, high quality contributions, and after~\citet{zhu2011identifying} that direct messages can be used to raise such engagement. Thus we might conclude that a simple network measure reflecting the number of communication links between project members (e.g. average node degree) would capture the extent to which such a network can transmit motivational messages. However, affective processes within a group are not the same as information transfer. For example, informing all group members about a task requires only one message per member. Not surprisingly, such processes on Wikipedia are automated – Wikiprojects use newsletter or mass-messaging to spread information broadly. However, raising motivation and building strong social ties requires reciprocity and perseverance; that is, multiple communication acts – messages – are conducive to a stronger relationship. Therefore, average volume of messages per node (i.e.~average node strength) – rather than average degree – would be a better measure of Wikiprojects’ capacity to build engagement.

\subsection{Complex coordination: Determinism, Degeneracy and Effective Information in networks}

Finally, following~\citet{nowakFunctionalSynchronizationEmergence2017} we assume that complex functions in social systems require complex coordination patterns between elements. Theoretical work by~\citet{tononi1998complexity} showed that structures that are capable of complex coordination need to have local specialization and global integration. Examples of such complex coordination can be found in brain imaging studies and neural modelling~\citet{tononi1994measure}. These concepts from information theory have recently been adapted into network measures of determinism (local specialization) and degeneracy (reverse of global integration)~\citep{kleinEmergenceInformativeHigher2020}. Both can be used to assess a social system’s capability to coordinate in a complex manner.

We follow~\citet{kleinEmergenceInformativeHigher2020} in our approach to quantifying structure and organization of the direct communication networks through the lenses of determinism, degeneracy and effective information. Here we review the main definitions.

A graph $G = (V, E)$ consists of a set of $n = |V|$ vertices and a set of $m = |E|$ links connecting selected pairs of vertices. Moreover, each link $(i, j) \in E$ has a weight $w(i, j) \in \mathbb{N}^+$ indicating the number of interactions between $i$ and $j$ and moreover $w(i, j) = w(j, i)$ for all $i, j = 1, \ldots, n$. Each graph can be represented by a $n \times n$ adjacency matrix $\mathbf{A}$ such that $a_{ij} = w(i, j) = a_{ji}$ if $(i, j) \in E$ and $0$ otherwise. Then, from any $\mathbf{A}$ a directed transition matrix $\mathbf{W}$ can be derived by dividing each row by the strength (sum of weights) of the corresponding node (rows with zeros only are left without change). This way values in each row are non-negative and sum up to $1$ so they can be interpreted as a valid probability distribution over possible next positions of a random walker starting at node $i$.

Then, determinism (local specialization of connections) of a graph G is defined as:
\begin{equation}\label{eq:det}
    \Det(G) = \log_2{n} - \frac{1}{n}\sum_{i=1}^n H(w_i)
\end{equation}
where $H(w_i) = \sum_j w_{ij}\log_2{w_{ij}}$ is Shannon entropy over the $i$-th row of the transition matrix $\mathbf{W}$ (isolated nodes are ignored in the calculation). Thus, determinism is the average certainty associated with the next move of a random walker placed randomly somewhere on the network. It is maximized when each node is connected to only one other node and minimized when each node connects to all other nodes. Hence, it is a measure of how specific local connections of nodes are on average.

Degeneracy of a graph G is defined quite similarly but in terms of Shannon entropy of the average over rows of W:
\begin{equation}\label{eq:deg}
    \Deg(G) = \log_2{n} - H\left(\frac{1}{n}\mathbf{1}_n^\top\mathbf{W}\right)
\end{equation}
where $\mathbf{1}_n$ is a vector of ones of length $n$ (again, isolated nodes are ignored). Thus, degeneracy is a measure of how concentrated the connectivity of $G$ is around a small subset of nodes. It is maximized when all nodes are linked to only a single hub and minimized when on average each node is visited equally often. In other words, it is a measure of the global integration of $G$. Example, artificial networks with representing structures of low and high  degeneracy are presented in Fig. 2.

\begin{figure}[!t]
    \centering
    \includegraphics[width=.495\textwidth]{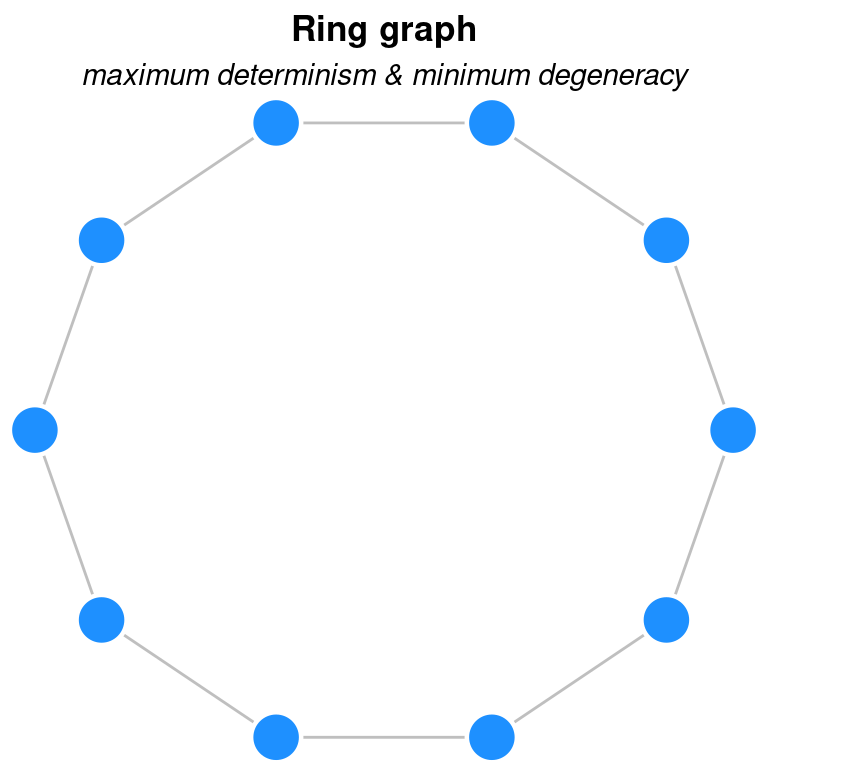}
    \hfill
    \includegraphics[width=.495\textwidth]{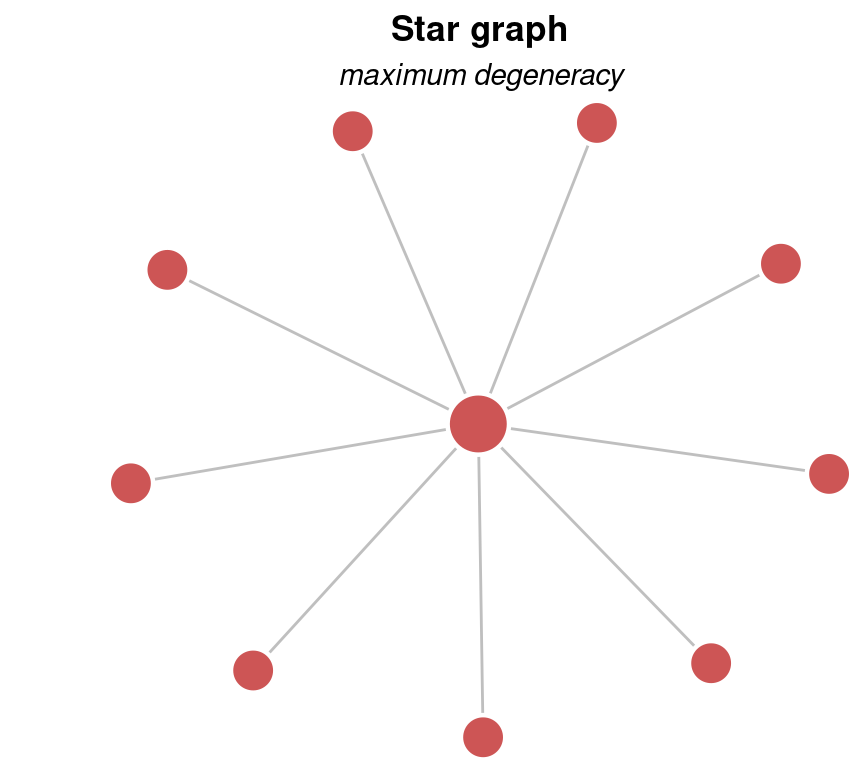}
    \caption{%
        Examples of networks with low and high measures of degeneracy; more examples can be found in~\citet{kleinEmergenceInformativeHigher2020}, figure 2.
    }
\end{figure}

Both $\Det$ and $\Deg$ have well-defined maximum values of $\log_2{n}$ so they can be normalized in the range $[0, 1]$ simply by dividing by $\log_2{n}$. Last but not least, they allow to define a quantity known as effective information which describes the overall structure of $G$ in terms of local specificity and global integration:
\begin{equation}\label{eq:ei}
    \EI(G) = \Det(G) - \Deg(G)
\end{equation}
The above definition of effective information mimics closely the theory of integrated information of~\citet{tononi1998complexity} and can be seen as a measure of non-trivial, complex structural organization in networks.

\subsection{Group size}

Acknowledging previous work on quality in peer production that focused on task processes within a group – e.g.~the composition of teams in terms of expertise or size of the collaborating teams~\citep{wilkinson2007cooperation} – we chose the size of the contributor group (i.e.~number of project members) as a control variable. By controlling for size of the projects we can assess whether affiliation processes contribute to group efficiency when controlling for task-related processes.

\subsection{Quality of group work}

The quality of group work that we aimed to relate to the properties of communication networks can be approximated by the quality tags assigned to articles by the Wikipedia community. The two ranks that are assessed by the whole community with respect to the same criteria are FA and GA (other ranks can have different criteria depending on the Wikiproject and the review process for them is carried out within projects). The numbers of FAs and GAs for each project are highly correlated ($r_{\text{Pearson}} = .92, p < .0001$) and thus they can be combined into a single measure of quality. However, projects differ vastly in their scope – i.e.~the number of articles that they curate – and we can expect that larger projects would naturally have a larger number of quality articles. For example, more articles within a thematic area can draw in more contributors interested in the various subtopics, who would self-select to work on the relevant articles. Moreover, the number of articles within a project’s scope is the cap on the number of possible quality articles. Thus an absolute value of quality articles will lead to a quality ranking with the top filled with projects of very large scope.

On the other hand, normalizing the quality score by the number of articles within a project’s scope (i.e.~the quality score becomes the fraction of high quality articles within all project’s articles) privileges smaller projects. In the most extreme case, a project curating only one article can achieve a hundred percent efficiency in producing quality articles with relatively little effort. Since the distribution of projects’ scope is highly skewed (most curate few articles), using such normalization would unduly lower the quality ranking of the largest projects.

We can thus consider a family of quality scores indexed by a parameter $p \in [0, 1]$ such that the quality score for a project $x$ is $Q_p(x) = N_Q(x)/n^p$  where $N_Q(x)$ is the number of FA+GA pages curated by $x$. Within this family the two cases described above constitute the extreme cases. A~quality score with no normalization by the projects’ scope is given by $Q_0$ and quality score normalized with the number of articles within scope is given by $Q_1$ . To~construct our response variable we have chosen the middle case in this family of quality rankings: $Q_{1/2}$ . In~effect, we normalize the number of quality articles not by the absolute number of articles within a project’s scope but rather by the size class of the project’s scope. Indeed, the square root of size serves as variance-stabilizing transformation and as such attenuates the effects of the highly heterogeneous distribution of project sizes while not discarding this information completely. An additional confirmation of the validity of this approach is given by the fact that $Q$-score with $p=1/2$ has almost perfectly log-normal distribution. Table~\ref{tab:descriptives} lists descriptive statistics for the chosen measures.

\begin{table}[!h]
    \centering
    \caption{Descriptive statistics of project measures}
    \begin{tabular}{lcc}
        Variable & Mean (SD) & Median  \\
        \hline
        Quality & 1.172 (1.684) & .652 \\
        Fraction in communication network & .512 (.173) & .503 \\
        Determinism & .761 (.093) & .782 \\
        Degeneracy & .265 (.139) & .238 \\
        Average connection strength & 30.094 (39.185) & 17.143 \\
        Number of project members & 136.39 (249.98) & 61 \\
        \hline
    \end{tabular}
    \label{tab:descriptives}
\end{table}

\subsection{Modelling quality}

To verify if group integration (fraction of project contributors within communication network), engagement building social ties (average node strength), and communication structure allowing for complex coordination (determinism and degeneracy) impact the quality of artifacts produced by peer production groups, we tested a linear regression model with all the above variables as predictors, quality score normalized by the square root of articles within scope as the response variable and number of editors listed within the project as control variable (Table~\ref{tab:model},~Model~1). Quality score, number of users within the project and average node strength were approximately log-normally distributed so we used their log transforms in the model. The model explained 23\% of variance in the quality scores of Wikiprojects ($F(5, 991) = 59.66, p < .001$). All predictors except the fraction of users within the communication network were significant. We tested a model (Table~\ref{tab:model},~Model~2) without the non-significant predictor, and the decrease in variance explained was not significant ($F(4, 991) = .178, p > .05$). The model without fraction of project members within the communication network also explained 23\% of the variance in quality scores ($F(4, 992) = 74.6, p < 0.001$).

\begin{table}[h]
\begin{threeparttable}
    \centering
    \caption{Effects of network structure on quality}
    \begin{tabular}{lccc}
        Predictor & Model 1 & Model 2 & Model 3  \\
        \hline
        Fraction in communication network & -.093 (.22) & & \\
        Determinism & 1.313$^*$ (.54) & 1.33$^*$ (.54) & \\
        Degeneracy & -.665$^*$ (.30) & -.71$^*$ (.28) & \\
        Average connection strength (log) & .285$^{***}$ (.04) & .278$^{***}$ (.04) & .246$^{***}$ (.03) \\
        Number of project members (log) & .315$^{***}$ (.04) & .32$^{***}$ (.04) & .347$^{***}$ (.03) \\
        Effective information &  & .667$^*$ (.28) \\
        Constant & -3.369$^{***}$ (.38) & -3.41$^{***}$ (.37) & -2.994$^{***}$ (.18) \\
        \hline
        R$^2$ & .231 & .231 & .23 \\
        \hline
    \end{tabular}
    \label{tab:model}
    \begin{tablenotes}
        \small
        \item Note: Linear regression coefficients; standard errors in parentheses; $^{***}: p \leq .001$, $^*: p \leq .05$
    \end{tablenotes}
\end{threeparttable}
\end{table}

Note that the regression coefficients for determinism and degeneracy are similar but with opposite signs. We can confirm this with a test of linear hypothesis: $H_0: b_{\text{deg}}+b_{\text{det}} = 0\ (F(2, 992) = 2.11, p > .05)$. The opposite impact of degeneracy (global integration) and determinism (local specialization) is indeed proposed to characterize complex systems. The measure of effective information, defined as $\EI(G) = \Det(G) – \Deg(G)$~\citep{kleinEmergenceInformativeHigher2020,tononi1998complexity} describes how much information is integrated within a complex system and can be used as an estimate of the system’s complexity. It has even been suggested as an approximation of the level of consciousness of a system~\citep{tononi2012integrated}.

\begin{figure}[!t]
    \centering
    \includegraphics[width=\textwidth]{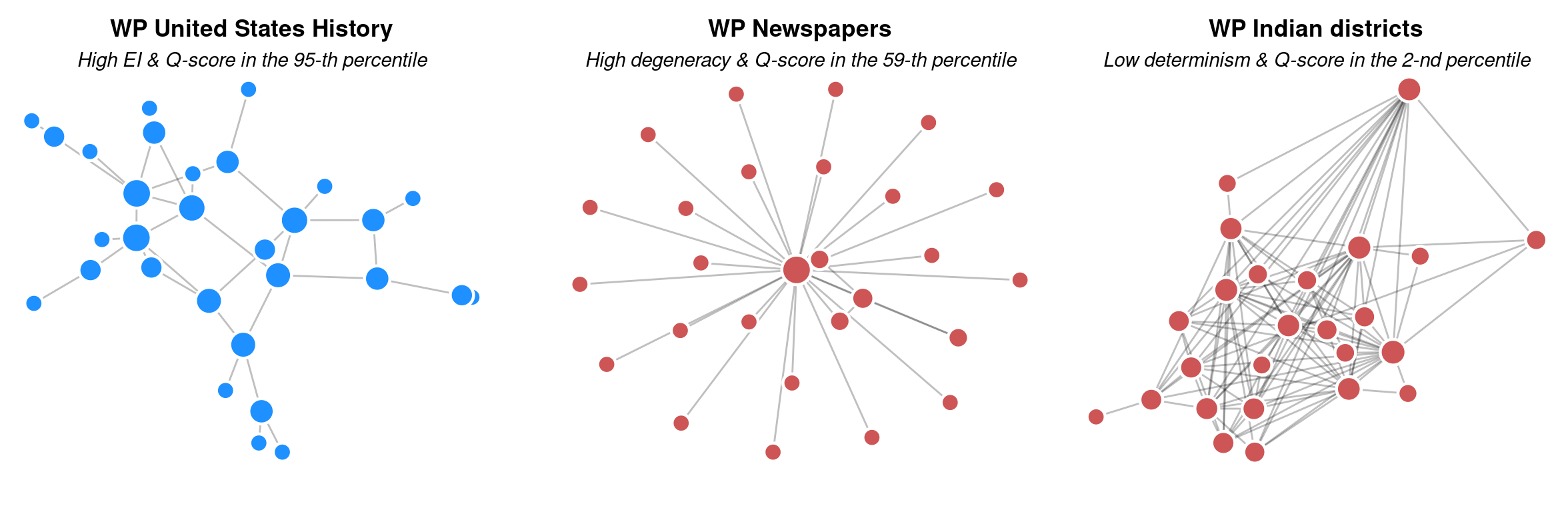}
    \caption{%
        Sample Wikiproject communication structures, varying in degeneracy, determinism and position in the quality ranking.
    }
    \label{fig:wp-networks}
\end{figure}

We tested a simplified model with degeneracy and determinism predictors replaced with effective information (Table~\ref{tab:model},~Model~3). All predictors were significant, the model explained 23\% of variance in quality scores $(F(3,993) = 98.65, p < .001)$, and in that was not significantly different from the model with degeneracy and determinism $(F(2, 992) = 2.11, p > .05)$. Example networks, depicting various levels of effective information are presented in Fig.~\ref{fig:wp-networks}. Thus we can conclude that effective information of the communication structure within Wikiprojects plays a significant role as a factor facilitating production of high-quality content on Wikipedia.

\section{Results – interviews}

\subsection{Analytical strategy}

Our quantitative analysis of the structure of one-on-one communication networks within Wikiprojects indicated that quality of their collective output grows with the strength of the ties between Wikiproject members, with the specialization of local connections (i.e.~when project members have specific partners with whom they communicate more than with others), and with integration of information among all group members (i.e.~when the connections in the network do not coincide on a limited set of nodes but rather the information spreads among all parts of the communication network). To further verify these findings and to gain an in-depth understanding of why such communication structures form and in what way they contribute to group collaboration, we interviewed selected Wikiproject members. We focused our analyses of this qualitative data on the uses of direct communication within group work and its role in fostering engagement and in coordinating group work. The choice of an established, highly successful Wikiproject provided us with information on how direct communication contributes to success. Wikiproject Tropical Cyclones ranked as the project with highest quality of collective output. The choice of a fast-growing project allowed us to understand what processes lead to the formation of particular patterns of communication. Wikiproject COVID-19 was started on 15.03.2020 and by February 2021 had accumulated 1375 articles and gathered 274 participants.

\subsection{The role of direct communication}

At the moment of founding, a new project’s communication is concentrated within the project’s coordination page and thus easily accessible to all prospective members. In the case of our analyzed COVID-19 project, the quick growth in interest and participation resulted in a fast flow of information: \enquote{You also just saw lots of people contributing to those initial discussions on the main project talk page. There were so many discussions being started and ideas being floated (\ldots) that it's almost overwhelming to follow.} (Interviewee 1)

This flood of discussions resulted in diversification of communication channels. More specific discussions (pertaining e.g.~to the epidemics development in specific regions) migrated, in an organic manner, to the talk pages related to specific encyclopedic articles: \enquote{a lot of the discussion that you saw early on on the project talk page has migrated to the article space} (I1). This left the project’s main coordination space as a place for more general discussions:  \enquote{the number of discussions being generated has of course declined a lot but it [is] still active. You still have editors asking thoughtful questions and soliciting feedback on specific pages.} and \enquote{\ldots I feel like it's still being used as a space for new information}. (I1) Interestingly, this early diversification of information points to an important role of direct messages posted on users’ talk pages.

Interviewee 1 said that he himself had little time to engage in interpersonal communication but that it was a common vehicle to provide feedback and corrections to the fast paced growth of contributions to the articles the project curated. \enquote{There were a lot of people who probably felt that their contributions to edit to Wikipedia were reverted [but] their additions were just relocated. (\ldots) Moving those requests to other spaces, I'm sure there was a lot of back and forth between editors on their talk page.} (I1) It seems that in this case direct messages were used as corrective feedback and explanation of actions taken by other editors. Importantly, \enquote{back and forth} exchanges took place, confirming our intuition that direct messages are used often not as a simple transfer of information but as a vehicle to build common understanding of the situation of group work: what are the accepted actions (norms) and what an article should look like (goals). In this, such messages might have reduced the discouragement of editors whose contributions were not accepted as they were but rather were moved to other articles.

Direct messages very often play the role of positive feedback as well. Such messages may take the form of a standardized text filled in with particular information (i.e.~a~filled in template), which can be pasted onto a user’s talk page using semi-automated tools such as the Twinkle Wikipedia add-on. A specific form of such templates are barnstars, acknowledgements and other awards which take the form of an image together with a personalized text message: \enquote{If someone writes a good article or like, they really show that initiative, they'll probably get a couple barnstars from a few different users.} (I3) More personalized feedback is delivered via messages written especially for the occasion. Interviewee 4: \enquote{I'd rather say thank you in my own words. I'd rather go and write, you know, thanks for your help with that. It was really useful.} In addition, the \enquote{thank you} function (an option available when viewing the edits contributed to a specific article) is an automated tool frequently used to give positive feedback, once an editor makes a good input (Interviewees 1 and 4). In contrast to personalized messages, such \enquote{thank yous} are not visible on the user talk page, but rather appear as a notification in the user profile. This gives editors a range of options to provide feedback that ranges from automated to very personalized, amplifying their capability to build social ties of different strength.

Negative feedback also ranges from more to less personal. Interviewee 4: \enquote{I also use twinkle to issue warnings and notifications. To people, I don't know. (\ldots) But for people I know I'm more likely to use a talk page message.} The templates available in the Twinkle tool allow users to provide depersonalized feedback and also to speed up the communication. This presumably allows to streamline task-related work. However, as the above quote shows, when there is already a relationship between users (a stronger social tie), a personalized message proves a better option: it might invite a reply and thus a deliberation on the shared understanding of norms and goals of the group. Thus, direct communication differs depending on the closeness of the interlocutors, with personalized messages inviting reciprocity and thus increasing the strength of the social tie.

The role of personalized, one-on-one communication is even more vivid when projects engage in soliciting new editors and in socializing them with the group. Communication on user talk pages is a way to establish sustainable relationships between more experienced users and newly joined members: \enquote{(\ldots) I tell them to leave me a talk page message [as a] part of their training. So again, they get the experience of doing that and I capture their username. So it's quite a cheating way of gathering everybody's username.} (I4). After leaving their name on each other's talk pages the editors are no longer anonymous strangers – they establish a personal, one-on-one channel of information exchange.

Experienced editors teach new project members about what needs to be edited and what makes a good edit. Interviewee 4 provides a detailed description of such training: \enquote{(\ldots) I train a lot of new people and I watch their early edits and I use welcome templates purely because it's a lot easier than trying to remember all the links that you want to give them for help pages and so on. (\ldots) And I always look for their first proper edits outside of their sandbox and leave a thank [you] with the thank [you] mechanism. Partly to thank them and encourage them but also so that they see how that works.}

Over time and with increasing volume of communication, relationships tend to transcend from task into more socio-emotional realm. This is accompanied by further diversification of communication channels. Interviewee 3: \enquote{A lot of it [is] user talk pages, there's a Facebook group among a couple users, we're all Facebook friends. (\ldots) there's an IRC Channel. There's also a Discord channel. So there's lots of different ways that the users will talk to each other. And it's become a bit of like socializing as well. Like there's some users I talk to almost every day. (\ldots) In a way, they've become almost a family of sorts.} Interviewee 4: \enquote{Off-Wiki friendships develop, and they obviously develop sometimes within projects. And you can tell people are making reference to things that they've talked about in email or that they met in a purple gun for a coffee or had a pizza or whatever}. Such strong, social bonds impact collaboration, for example: \enquote{I'm probably more likely to help if it's somebody I've met and work[ed] with previously than if it's just a name on a talk page, I'm not saying I wouldn't help and I'm not saying it's the only reason that I would help, but it's certainly encourages me to get involved.} (I4)

In sum, throughout a project's existence, collaboration is coordinated via various channels of communication that serve different communicative purposes. Direct messages on users’ talk pages often serve to invite and socialize new members and to provide feedback to encourage further contributions. Stronger relationships are enacted through more personalized messages, presumably to invite reciprocal and more prolonged exchanges.

\subsection{Organization and coordination of group work}

Interviewee 1 defines Wikiproject as \enquote{a single space where lots of editors (\ldots) come together and really work to bring some consistency across} thus underscoring the importance of coordination of work within the group. Good projects differ from bad projects in the extent to which they are able to maintain \enquote{an active community (\ldots), engagement with the wider community and an outreach} by organizing and announcing editathons, campaigns and competitions (I4). Thus, the distinction between successful and unsuccessful projects has a lot to do with how the project is connected to the wider Wikipedia community: \enquote{Good projects (\ldots) interact with the rest of the community to help people to understand the technicalities, to reach out to people who are working in that sort of area and involve them. And the bad projects see themselves as being the custodians, the ring fences, the owners of stuff} (I4). From this we can understand that good projects are able to tap on resources available in the community, while bad projects remain a specialized, closed group of contributors.

This understanding of success as an ability to mobilize distributed resources into a single effort is also reflected in how the work within projects is organized. When asked about division of labor and coordination of tasks, the interviewees underscore the bottom-up, organic manner in which such processes take place. In the newly formed COVID-19 project editors try to \enquote{find their niche, their dedicated space within the project (\ldots), saying: I want to focus on this or I'm gonna sign up to focus on [COVID-19] case counts in Italy or Spain (\ldots)} (I1). Similarly, our interviewee from the Tropical Cyclone projects reminiscens:  \enquote{I~remember these users (\ldots) just took the initiative to make standardized info boxes, eventually getting standardized [hurricane] track maps, making sure that all the different hurricane articles look pretty much the same, that they were high quality, that they were referenced, that they weren’t using blogs for information.} (I3) In a sense, \enquote{the division of labor (\ldots) kind of became self-evident} (I3).

At the same time they notice that there is little in the way of formal roles or leaders that would serve as coordinators: \enquote{(\ldots) People just wanted to help organize the movement's efforts, (\ldots) so that editors weren't just thoroughly confused about where they were supposed to be adding new information. I wish I had a more clear answer for you on leadership. (\ldots) I'm sure there were people who stepped up and, you know, really took initiative.} (I1) \enquote{I haven't seen any sort of more formal structure. The only [projects] that tend to have a more formal structure are the ones where there is a user group off Wiki. So Wikiproject Medicine has the medicine foundation with a board and fundraising (\ldots)} (I4). In effect, the within project coordination becomes seamless and implicit, as users self-select to fill in where others left work to be done: \enquote{It's almost an informal collaboration, like we know what's expected so it's not exactly coordination} (I3).

In the case of established projects, this smooth collaboration might be an effect of experience, of editors internalizing their specific roles in the group effort. \enquote{Over [the] years we've collaborated and discussed (\ldots) and in the new [hurricane] season articles (\ldots) people can look back at other years and realize oh that's what we've done, so that's how it happens (\ldots).} (I3) It’s an effect of \enquote{past years’ work, coordination, reviewing the good articles, setting the standards, making sure that all the articles look the same}. (I3) The project works like well-oiled machinery, where the members are specialized in their tasks but at the same time rely on others to fulfill theirs.

Communication plays an important role in making such a system work. Project members specialized in certain areas (e.g.~making maps for hurricane tracking) naturally communicate with each other, but they also need to reach out farther - within and across projects - to tap on resources and skills they do not possess. Interviewee 4: \enquote{I go to the project page and find somebody, and then go and talk on their talk page. (\ldots) If I need to go and ask for some advice or get a second opinion (\ldots), you might have a source that I want to look something up for me.} Whilst direct messaging might be one of the choices for such advice seeking, other means of communication may be used. \enquote{I will private message them or leave them a message on their talk page. And in one or two cases, contact them off Wiki:  a couple of them I know by email, or by Facebook. And I might just say, do you have this paper? Or I'm trying to find out this person's date of birth. Can you look it up in one of your sources for me?} (I4) 

Such resource seeking might also happen on a project’s page to increase the visibility of the request or in situations when the editor does not have a personal relationship with anyone with fitting expertise. Interviewee 4: \enquote{you use your social capital, to use the jargon, to get help and to help other people. (\ldots) There are a few Wikiprojects that I go to when I have a question. (\ldots) If ever I'm writing a biography of somebody and they have some experience in the military (\ldots), the Military History project is a good place to go and check. (\ldots) Ask people to look things up in resources that they may have subscriptions to. (\ldots) Make sure I've got the terminology correct, things like that. (\ldots) There's a UK Railway project or rather a route and international railway project and that's simply a good place where you can go and get technical information and get people to look stuff up. Being sources for you.}

Project members also notify each other when they notice something that might be of interest to somebody. Interviewee 4: \enquote{I also tend to drop things into projects (\ldots) if I think they'll be of interest. So if I see something in the news and I can't write about it, but I think somebody might be interested and I know it's an active project in an area that I'm interested in, then I'll go and leave something on a talk page.}

Thus it seems members of Wikiprojects treat each other as a reliable source of information. Interviewee 2: \enquote{I've got access to Discord Channel, I've got access to the IRC, I can just generally ask: can you just take a look at this article? See what you think? [Am I going ] in the right direction. Am I wrong?} However, information exchange relies on previous experience with a particular Wikipedian and mutual trust. Interviewee 4: \enquote{If you (\ldots) use the [project] talk page to ask a question, you know which group of people will be replying and indeed, you know which ones you can rely on if you get an answer. (\ldots) You think: well, I've interacted with them often enough and seen their interaction with other people. Often up, I know they have a level of expertise and they're reliable and they [are] trustworthy in terms of the technical response. And indeed more generally.}

In sum, Wikiproject’s divide and integrate their work through a bottom-up self-selection to tasks. Experience in collaboration helps them develop procedures for implicit coordination of tasks and allows them to build strong ties that result in trust and reliance on each other. Moreover, project members specialize in specific tasks and learn where to seek advice and help when they need information or skills beyond their own expertise. Communication within and across projects helps to connect such specialists; however trust is crucial to rely on such sources. In this, Wikiproject’s display a high level of social capital: even if project members do not know something, they proverbially \enquote{know who knows} and are able - through their stronger (personal relationship) or weaker (through projects or initiatives they know) ties - to tap on this knowledge.

\subsection{Summary of results}

Our quantitative investigation into Wikiprojects’ direct communication structure showed that strength of the relationships within the group - as measured by the average number of messages sent over the social ties - correlates with higher quality products of collaboration. By interviewing selected members of an established and a growing Wikiproject we confirmed that direct messages can serve as a channel for socializing new project members and for sending positive and corrective feedback. Other popular means of communication within Wikipedia - such as project talk pages - are used for setting the agenda for future work or soliciting less personal feedback from a broader group of project members.

Moreover, our interviewees differentiated between automated and personalized messages: those framed as a standardized warning or acknowledgment of work were used for contacting distant or altogether unknown relations, while personalized messages were written for the strong and close connections. Thus, not only the volume of communication but also the form of messages can convey how close the interlocutors are. In sum, personalized, direct communication seems to complement more official, group communication channels by providing a vehicle to build strong social ties that facilitate affective processes within the group. In contrast, the fraction of group members that take part in direct communication did not play a part in group efficiency. We may conclude that what matters is the quality of relationships, not their span.

The other important result from the network analysis is that successful Wikiprojects have a direct communication structure that allows for complex coordination: for sharing of information locally, between project members clustered into small groups, and at the same time globally across such groups rather than by a single coordination center. In interviews, Wikiproject participants shared that they preferably seek advice from and offer help to those whom they trust and whose expertise they are aware of; thus they operate within a close circle of strong social ties. At the same time, they are skilled at using group level communication (e.g.~project talk pages) to acquire information that their closest relations do not possess. In effect, whilst locally they might communicate with a restricted number of collaborators, globally, it is easy for any resource or information to traverse the network. Thus both quantitative and qualitative analyses converge to show that complex patterns of information integration and coordination within Wikiprojects are related to better group performance.

\section{Discussion}

In our studies we attempted to verify how affiliation processes (affective domain) complement task-related coordination (instrumental domain) within groups collaborating online in the course of producing high quality output of collective effort. As argued by early influential research on group interaction~\citep{balesEquilibriumProblemSmall1953,balesHowPeopleInteract1955}, group dynamics must be studied in the context of these two conflicting developmental tendencies within the group, as they create inherent challenges that determine how the group will operate~\citep{balesAnalysisSmallGroup1950,balesPhasesGroupProblemsolving1951}.

The first type of challenges pertains strictly to the group task and how effective the group is in its achievement (instrumental domain). The second type of challenges relates to the extent to which group members are able to maintain close social ties and sustain group cohesion (expressive domain). Tensions caused by the instrumental challenges are reduced through division of resources and labor, and establishment of status and authority hierarchies within the group. On the other hand, tensions within the expressive domain induce behaviors that are focused on integration of opinions, and uniformity among group members. Moreover, according to~\citet{mcgrathGroupsInteractionPerformance1984}, affective processes have a regulatory (reinforcing and directing) function in relation to task behavior. Thus, a group is a system in a dynamical equilibrium between the conflicting drives for cohesion and differentiation~\cite{balesEquilibriumProblemSmall1953}. With the growing ubiquity of virtual collaboration - a tendency which the pandemics has strengthened and possibly calcified - it is important to understand how this balance can be achieved even when groups collaborate solely via new media. Wikipedia is a renowned success case of such technology mediated collaboration.

A growing body of work investigates task related coordination on Wikipedia. An influential study has found that implicit coordination - in the form of division of contributors into a highly productive core and periphery - was related to higher quality of articles, more so than the number of editors or explicit coordination in the form of discussions on the article talk pages~\citep{kitturHarnessingWisdomCrowds2008}. Other studies, however, found that numbers matter: both a higher number of editors, as well as a higher number of edits correlated with high quality~\citep{wilkinson2007cooperation}. An explanation of this discrepancy might be that it is important who comprises such a high number of editors. For example, varied background and diversified experience in editing~\citep{sydow2017diversity}, longer tenure, and willingness to perform multiple roles (functional tasks within article editing,~\citet{liuWhoDoesWhat2011}) positively impact quality of the articles on which editors work.

Explicit coordination has also been proven to matter. The structure of communication on Wikiproject talk pages impacts projects’ efficiency (measured as number of edits to articles): again, existence of a core of editors with high numbers of discussion posts and high values of betweenness centrality in the discussion network, was beneficial. Thus both implicit (division into core and periphery) as well as explicit (discussions) task related coordination is correlated with higher efficiency in pursuing collective goals. There is much less known, however, about the affiliation processes in online collaborative environments. Our study fills this gap.

First, interviews’ analysis showed that Wikipedians diversify their communication, depending on the communicative goal (task or affiliation related) and strength of relationship. The preferred task-related coordination space are talk pages - both of encyclopedic articles and of Wikiprojects. The affective processes, on the other hand, take place on user talk pages where Wikipedians post direct messages to one another, or outside of Wikipedia, on various social media sites. While such direct messages can have multiple goals (e.g.~they can also convey task related information) they are a dominant place to send general feedback (both positive and negative) which may help sustain engagement of contributors as well as socialize them with the collaborating group. Feedback messages were previously identified as types of leadership behaviors on Wikipedia - transactional leadership (positive feedback) and aversive leadership (negative feedback)~\citep{zhu2011identifying}. Our results enhance these findings by showing that feedback is preferentially provided via direct messaging.

Moreover, our quantitative network analysis also confirmed that direct communication impacts the effects of group work: high average volume of direct messages was correlated with higher quality of Wikiprojects’ output. Thus our results are in line with the so far scarce studies on effects of direct messaging on Wikipedia’s collaboration: that articles benefit from their contributors engaging in private messaging in addition to discussions on articles~\citep{parkCommunicationBehaviorOnline2016}, and that direct communication structure that allows linkages across and within contributors from core and periphery relates to higher quality of project output~\citep{rychwalskaQualityPeerProduction2020}. The results presented here go beyond these previous findings by showing that high intensity of affective processes is beneficial for effectiveness of online group collaboration. This result, combined with previous studies on task related coordination, shows that online groups, just like their offline counterparts, need to find a balance between affiliation and instrumental processes within the group. Understanding the interplay between affective and instrumental processes is important given that even on a such task focused community as Wikipedia, personal likes and dislikes may drive editing dynamics in a negative way (e.g.~into edit warring), wherein affiliation with particular others may be more important than the quality of contributions~\citep{lernerFreeEncyclopediaThat2020}.

Arguably, the most important and novel contribution of our paper, however, is that direct communication structure that allows for complex information integration within the group is related to higher quality of articles curated by Wikiprojects. That is, in successful projects, a particular individual’s contacts are relatively deterministic: she has a select group of relationships with whom she tends to communicate much more often than with others. On the other hand, in these successful projects, there is little determinism on the global, group-wide scale: if a particular piece of information or a resource is acquired by any member of the group, the next steps of its way across the group are on average quite unpredictable. In this way the effective mode of operation is that, locally, information is processed within circles of individuals and globally, there are sufficient connections distributed over the entire network to allow the information to spread rather than to be captured by a limited set of individuals. The interviews further elaborate on this result from network analysis: project members build strong trust and expertise based ties by history of previous interactions, but they also have multiple connections to others with different expertise through weaker relations. Such weak ties can exist within and across projects, as evidenced by our network analysis and interviews, respectively.

Our finding that global information integration (lack of a well determined sink for information or a finite coordination center) is beneficial for group effectiveness does not preclude the existence of a core of project contributors. Indeed, the distributions of both the number of messages sent as well as of degree centrality of Wikiprojects’ members tend to be highly skewed and heavy-tailed, suggesting that there are members who communicate more intensively than others and whom we might call “core”. What our results suggest, however, is that in online collaboration groups the core should not strive to be a pipeline for information and resource spread. 

Specifically, globally integrated dynamics might be beneficial for more complex tasks, such as upgrading an article to a high quality rank. \citet{qinInfluenceNetworkStructures2015} have shown that the number of edits to articles within a projects’ scope is positively related to existence of leadership behaviors such as dominance in project talk page discussions (measured by betweenness centrality), which might suggest that centralized communication is favorable for Wikiprojects. However, mere adding of edits is a much simpler task than promoting an article to featured status. Indeed, \citet{nemotoSocialCapitalIncreases2011} have shown that promotion of articles to lower ranks of quality requires more centralized leadership (as measured by group degree centrality of direct communication networks) with less cohesive connections (measured by average clustering coefficient) than promotion to FA status. In this most complex task, cohesiveness was more important than centralized leadership. Our results show that this difference might be due to a larger press on global integration of information in such complex tasks. The practical implication of this finding is that in management of online collaboration, project members who are central in communication should strive to maintain redundant ties with the periphery, and encourage links across the periphery, so that the core of the group does not become a bottleneck in information integration or resource mobilization. 

Such redundant connectivity across specialized local groups might be important specifically in online collaboration environments which are often characterized by high fluidity in membership. In standard organizations, maintenance of a skilled workforce can be achieved by, e.g.~financial incentives. Such solutions are not available in volunteer-based peer production and this often results in high turnover of contributors. For example, on Wikipedia, contributor career paths are volatile and unpredictable: editors switch between various functions and roles as they gain experience and self-select to favorite tasks. Some might take a long \enquote{Wikibreak} or even leave the community for good. At the same time the global proportions of functional roles are quite stable, implying that when individuals abandon certain roles, others take their place spontaneously~\citep{arazyTurbulentStabilityEmergent2016}. Our interviewees confirm that role uptake is organic and spontaneous. In such a working environment, no person in the direct communication network should be irreplaceable. Redundancy, not only in local, usually dense, ties but also in those weak, wide-spanning connections is critical to maintain sufficient levels of coordination in Wikiprojects, given the possibly fluctuating membership. In effect, our results suggest that complex tasks undertaken by online collaborating teams are benefited by communication structures that are characterized by specificity in local connections together with wide-reaching, integrative global links. 

Such connectivity structure - and indeed the measures for effective information - might be related to high social capital of a group: \enquote{the sum of actual and potential resources embedded within, available through, and derived from the network of relationships possessed by an individual or social unit}~\citep[p.~243]{nahapietSocialCapitalIntellectual1998}. Social capital of a group is often defined in terms of bonding - i.e.~the cohesiveness of ties within a network. In offline teams an established network of close relationships within a social system increases commitment~\citep{brassBeingRightPlace1984,colemanFoundationsSocialTheory1994} and a higher density of connections leads to a higher likelihood of collective endeavors~\citep{putnamTuningTuningOut1995}. On Wikipedia, such social capital brought to an article and accumulated by previous collaboration, results in higher article quality~\citep{nemotoSocialCapitalIncreases2011}. Strong connections in a network of co-contributions to sentences within an article - also a measure of bonding capital - are also related to higher quality~\citep{liuWhoDoesWhat2011}. However, such capital is different from bridging capital of individuals that can be accrued by occupying a structural hole~\citep{jacksonTypologySocialCapital2020}.

Bridging capital is highest for individuals when they are unique intermediaries between parts of a network~\citep{jacksonTypologySocialCapital2020}. In offline studies on social capital an important role in communication networks is played by the so-called gatekeepers - a small number of key group members with expertise of how to find relevant information within a system~\citep{su2011multidimensional}. These gatekeepers are crucial for the group to work efficiently and that crucial position gives them power and control. Yet, for the group their uniqueness might be detrimental - they are a bottleneck in group coordination. As discussed above, in online systems that attempt complex tasks requiring specialized knowledge or skills, and which are characterized by high fluidity in membership, gatekeepers could prove disastrous: if they reduce their activity or simply are overwhelmed by amount of work, the whole group task (i.e.~FA~candidate article) might fail. Our results suggest that successful online teams are characterized by a communication structure that lacks any such gatekeepers. Future research might investigate whether complexity of the task (i.e.~comparison between different quality grades of articles), size of the team, as well as fluctuations in membership impact what communication structures are effective for an online peer production group.

\section{Limitations}

Without a doubt Wikipedia is one of the greatest success stories of collaborative knowledge production with no tangible financial incentives. As such it should be studied as this can reveal factors responsible for its unique growth and success. However, for the very same reasons our results should not be generalized to other similar platforms without additional considerations. Wikipedia may differ in important and unique ways from an average collaborative peer-production platform which can affect the scope of generalizability of our results. Studies similar to ours but based on data from other platforms need to be conducted in order to assess this.

Furthermore, in this study we focused exclusively on the structure of direct communication networks. It may be of interest to replicate similar analyses for other communication layers in the Wikipedia ecosystem, in particular discussions on article talk pages in the so-called main namespace as well as those on pages of individual Wikiprojects (project namespace). Moreover, correlations between communication structures at these different layers may provide additional insights and could be studied with methods developed for multilayer networks~\citep{boccalettiComplexNetworksStructure2006a}.

Last but not least, we used time-aggregated networks and even though this simplification was still enough to reveal several structural properties linked to quality of output of Wikiprojects it did not capture the full richness of our data. Therefore, it may be worthwhile to also study dynamic properties of communication within Wikiproject. However, one should also note that for many smaller projects available data may be too sparse to allow reliable dynamic analyses leading to significantly smaller sample sizes than the one we considered here.

\section{Acknowledgments}

This work was supported by the Polish National Science Centre through grant 2017/27/B/HS6/00626.

\printbibliography

\end{document}